\documentclass[prd,aps,nofootinbib,preprintnumbers]{revtex4}

\usepackage{graphicx}

\begin{document}

\title{Quadratic chaotic inflation with a logarithmic-corrected mass}

\author{Shinta Kasuya and Mayuko Taira}

\affiliation{
Department of Mathematics and Physics,
     Kanagawa University, Kanagawa 259-1293, Japan}
     
\date{July 31, 2018}

\begin{abstract}
We consider a simple modification of quadratic chaotic inflation. We add
a logarithmic correction to the mass term, and find that this model can be consistent 
with the latest cosmological observations such as the Planck 2018 data, in combination with 
the BICEP2/Keck Array and the baryon acoustic oscillation data. Since the model predicts
the lower limit for the tensor-to-scalar ratio $r$ for the present allowed values of the 
spectral index $n_s$, it could be tested by the cosmic microwave background polarization 
observation in the near future. In addition, we consider higher-order logarithmic corrections. 
Interestingly, we observe that the scalar spectral index $n_s$ and 
$r$ stay in rather a narrow region of the parameter space. 
Moreover, they reside in a completely different region from that for the logarithmic
corrections to the quartic coupling. Therefore, future observations may distinguish which 
kind of corrections should be included, or even single out the form of the interactions.
\end{abstract}

\maketitle

\section{Introduction} 
Inflation is an interesting paradigm in the early Universe. It solves the problems in the hot 
big bang Universe and may also provide the seeds of the density perturbations
of the Universe. The simplest inflation model with a scalar field $\phi$ would be 
chaotic inflation \cite{Linde} with a quadratic potential,
\begin{equation}
V=\frac{1}{2} m^2 \phi^2,
\end{equation}
where $m$ is the mass of the scalar field $\phi$. However, this simple model is likely be 
excluded by recent observations such as the cosmic microwave background (CMB) observations 
by the Planck satellite \cite{Planck}. The main obstacle is that the tensor perturbations it predicts 
are too large because of the large potential energy at the large field amplitudes during inflation.

One way out is to somehow lower the potential at large field values. A lot of scenarios
along this line have been proposed \cite{rad1,rad2,rad3,poly1,poly2,poly3,nonK1,nonK2,nonK3,
nonK4,nonG1,nonG2,nonG3,nonG4,nonG5,nonG6,otherfield1,otherfield2,R1,R2,R3}. 
An interesting possibility is due to radiative corrections \cite{rad1,rad2,rad3}, 
where, in most cases, the quadratic potential becomes flatter because of the running 
of the quartic coupling of the inflaton.

In this article, we instead consider a very simple case where the mass has a running of the 
following form: 
\begin{equation}
m^2(\phi)=m^2\left[1-K\log\frac{\phi^2}{M^2}\right],
\label{mlog1st}
\end{equation}
where $K$ is a positive constant and $M$ is some large mass scale. 
Positive $K$ can be realized, for example, when couplings of $\phi$ to fermions dominate 
over those to bosons if the logarithmic correction is radiatively produced. As shown below,
we find that the model is consistent with the Planck 2018 data, in combination with 
the BICEP2/Keck Array (BK14) and the baryon acoustic oscillation (BAO) data \cite{Planck}.  
This is contrasted with the corrections due to the running of the quartic coupling, which 
may now be inconsistent with the combination of the Planck 2018 and BK14/BAO data.

In addition, we go further to include higher-order logarithmic corrections to the mass term.
It is interesting to note that the scalar spectral index $n_s$ of the curvature perturbation and 
the tensor-to-scalar ratio $r$ stay in a narrow region of the parameter space  ($n_s$, $r$), 
so that we could still find $n_s$ and $r$ consistent with the data
for slightly different model parameters. For comparison, we investigate higher-order 
logarithmic corrections to a quartic coupling of the potential, which places $n_s$ and $r$ in a
completely different region of the parameter space ($n_s$, $r$).

Notice that Ref.~\cite{rad3} considered the radiative corrections to the mass, including the 
second order of the logarithmic corrections to make a plateau in the potential. We do not consider 
such particular potentials in this article.

\section{Logarithmic-corrected mass inflation}
As mentioned, we consider a simple correction to the mass in quadratic chaotic inflation,
where the potential is given by \cite{rad3} 
\begin{equation}
V(\phi) = \frac{1}{2}m^2\left(1-K\log\frac{\phi^2}{M^2}\right)\phi^2.
\label{potmass1st}
\end{equation}
$M$ can be arbitrary, since the form of the potential does not change when we reparametrize 
as $M \rightarrow M'$, $K\rightarrow K'=K/[1-K\log(M'^2/M^2)]$,
and $m^2 \rightarrow m'^2=m^2[1-K\log(M'^2/M^2)]$. Thus, we set $M=M_{\rm P}$ below without 
loss of generality, where $M_{\rm P} (=2.4 \times 10^{18}$ GeV) is the reduced Planck scale.
We also assume that nonrenormalizable higher-order terms, for example, bound the potential 
from below at larger amplitudes, while they do not affect the dynamics of the 
inflaton during inflation.

Using the slow-roll parameters
\begin{equation}
\epsilon(\phi) \equiv \frac{1}{2M_{\rm P}^2}\left(\frac{V'(\phi)}{V(\phi)}\right)^2, \quad
\eta(\phi) \equiv \frac{1}{M_{\rm P}^2} \frac{V''(\phi)}{V(\phi)},
\end{equation}
where a prime denotes a derivative with respective to $\phi$,  
the scalar spectral index $n_s$ of the curvature perturbation and 
the tensor-to-scalar ratio $r$ are, respectively, 
\begin{eqnarray}
n_s(\phi) & = & 1-6\epsilon(\phi)+2\eta(\phi), \\
r(\phi) & = & 16\epsilon(\phi).
\end{eqnarray}
They should be evaluated at $\phi=\phi_N$, the field amplitude $N$ $e$-folds before 
the end of inflation. $\phi_N$ is related to $N$ by
\begin{equation}
N = \int_{\phi_{\rm e}}^{\phi_N} \frac{1}{M_{\rm P}^2}\frac{V(\phi)}{V'(\phi)}d\phi,
\end{equation}
where $\phi_{\rm e}$ stands for the amplitude of the field at the end of inflation, defined by
$|\eta(\phi_{\rm e})|=1$.

\begin{figure}[ht!]
\hspace*{12mm}
\includegraphics[width=100mm]{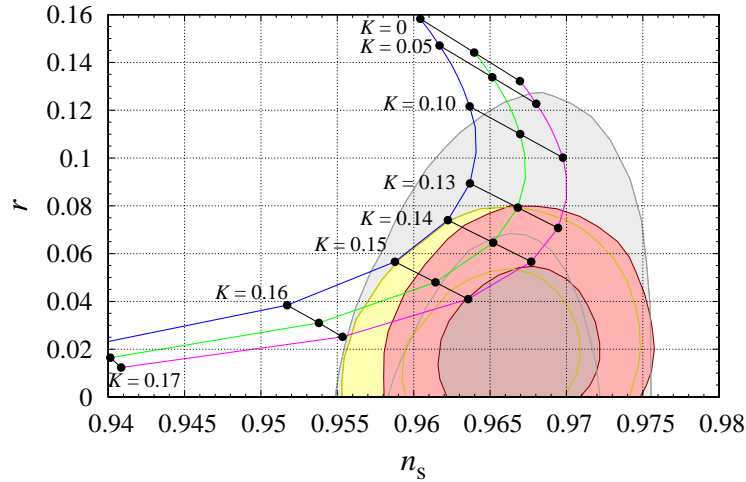} 
\vspace{13mm}
\caption{Spectral index $n_s$ and tensor-to-scalar ratio $r$ for $K=0-0.17$. 
The top (blue), middle (green), and bottom (magenta) lines are for $N=50$, 55, and 60, respectively. 
Notice that the $K=0$ case corresponds to the original quadratic chaotic inflation. 
The Planck-allowed regions are shown in dark (1$\sigma$) and light (2$\sigma$),
where we adopt the {\it Planck} TT,TE,EE+lowE+lensing (gray),  
Planck TT,TE,EE+lowE+lensing+BK14 (yellow), and
Planck TT,TE,EE+lowE+lensing+BK14+BAO data (red) \cite{Planck}.
\label{fig1}}
\end{figure}

We numerically calculate $\phi_N$ for fixed $N$ to obtain $n_s(\phi_N)$ and $r(\phi_N)$
for various values of $K$. The results are shown in Table~\ref{table1} and Fig.~\ref{fig1}. 
We also plot the 1$\sigma$ and 2$\sigma$ region allowed by Planck results \cite{Planck} 
in Fig.~\ref{fig1}.

\begin{table}[ht!]
\caption{$n_s$ and $r$ for some values of $K$.
\label{table1}}
\begin{tabular}{|c|c|c|c|c|c|}
\hline
\hspace{2mm} $K$ \hspace{2mm} & \hspace{1mm} $\phi_{\rm e}/M_{\rm P}$ \hspace{1mm} 
& \hspace{2mm} $N$ \hspace{2mm} & \hspace{1mm} $\phi_N/M_{\rm P}$ \hspace{1mm} 
& \hspace{4mm} $n_s$ \hspace{4mm} & \hspace{4mm} $r$ \hspace{4mm} \\
\hline\hline
& & 50 & 12.90 & 0.964 & 0.122 \\
0.10 & 1.175 & 55 & 13.50 & 0.967 & 0.110 \\
& & 60 & 14.08 & 0.970 & 0.100 \\
\hline
& & 50 & 11.98 & 0.964 & 0.089 \\
0.13 & 1.096 & 55 & 12.49 & 0.967 & 0.079 \\
& & 60 & 12.98 & 0.969 & 0.071 \\
\hline
& & 50 & 11.55 & 0.962 & 0.074 \\
0.14 & 1.070 & 55 & 12.01 & 0.965 & 0.067 \\
& & 60 & 12.45 & 0.968 & 0.057 \\
\hline
& & 50 & 11.03 & 0.959 & 0.057 \\
0.15 & 1.043 & 55 & 11.43 & 0.961 & 0.048 \\
& & 60 & 11.80 & 0.964 & 0.041 \\
\hline
& & 50 & 10.41 & 0.953 & 0.038 \\
0.16 & 1.017 & 55 & 10.74 & 0.954 & 0.031 \\
& & 60 & 11.03 & 0.955 & 0.025 \\
\hline
\end{tabular}
\end{table}

We can see that the tensor-to-scalar ration $r$ becomes smaller for larger $K$,
since the logarithmic term lowers the potential at large amplitudes. For an appropriate range
of $K$, the model is consistent with observations. In this model, it is most
favorable for $K\simeq 0.14 -0.15$ and $N\simeq 55 - 60$, which resides inside the 1$\sigma$
allowed region for the Planck TT,TE,EE+lowE+lensing data \cite{Planck}, 
and is even consistent with the Planck TT,TE,EE+lowE+lensing+BK14+BAO data \cite{Planck} 
for $K\simeq 0.14 -0.15$ and $N\simeq 60$.

It seems that the tensor-to-scalar ratio $r$ has a lower bound $r\gtrsim 0.02$
for a spectral index $n_s$ that is consistent with the Planck observations. Therefore,
this model could be tested by the CMB polarization observations in the near future.

\section{Higher-order corrections}
\begin{figure}[ht!]
\hspace*{9mm}
\begin{tabular}{cc}
\includegraphics[width=90mm]{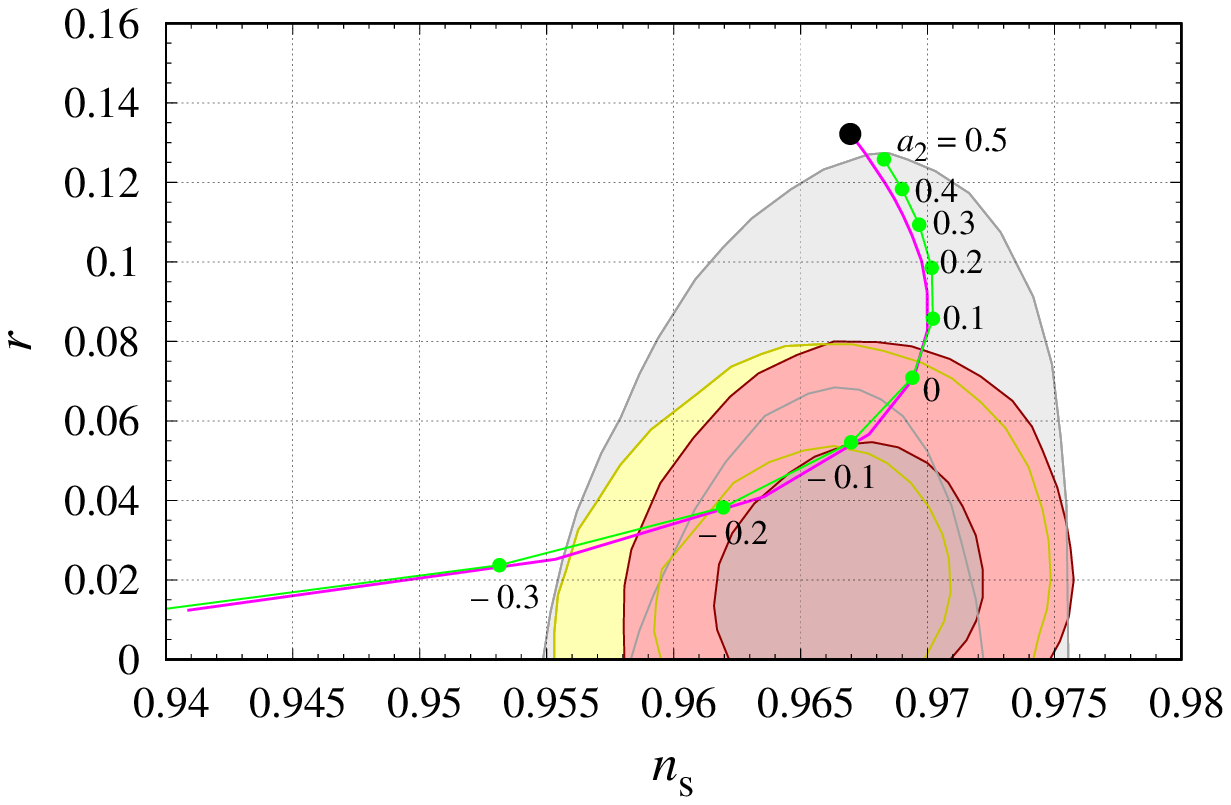} &
\includegraphics[width=90mm]{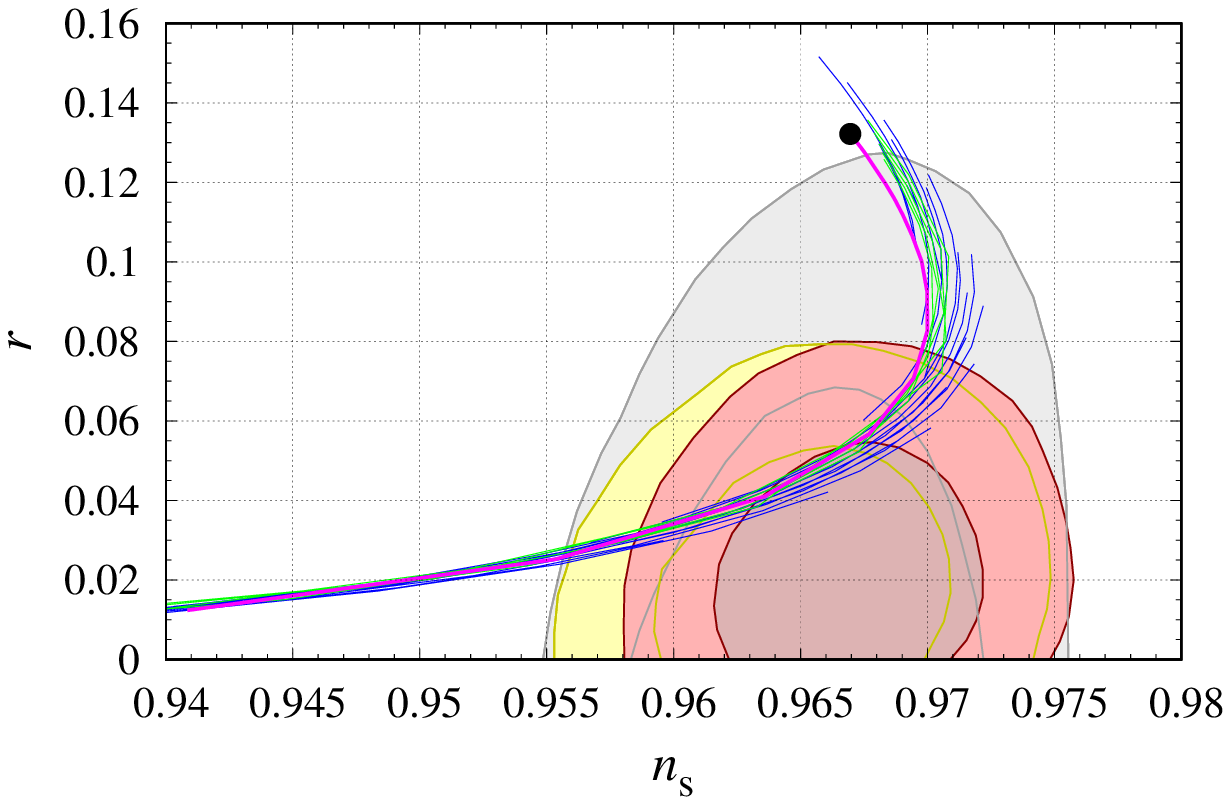} 
\end{tabular}
\vspace{13mm}
\caption{Spectral index $n_s$ and tensor-to-scalar ratio $r$ for $N=60$ in the case of
the higher-order logarithmic corrections to the mass. The magenta, green, and blue lines 
represent the cases with corrections up to first, second, and third order, respectively.
The black dot denotes the original quadratic chaotic inflation case. 
The Planck-allowed regions \cite{Planck} are the same as in Fig.~\ref{fig1}.
Left: The case for $a_2=(-0.4) - 0.5$, $K=0.13$ and $a_n=0$ $(n\ge3)$, as an example. 
Right: All the cases  
explained in the text.
\label{fig2}}
\end{figure}

So far we have considered a logarithmic correction to the mass term in the lowest order 
as in Eq.(\ref{mlog1st}). Here we investigate the effects of including higher-order corrections.
To be specific, we expand $m(\phi)$ in powers of log$\phi^2$, and
assume the following form for the logarithmic corrections to the mass term \cite{rad3}:
\begin{equation}
V(\phi)=\frac{1}{2}m^2(\phi)\phi^2=m^2\left[1-K\log\frac{\phi^2}{M^2} 
+ \sum_{n\ge 2}  a_n \left(K\log\frac{\phi^2}{M^2} \right)^n\right]\phi^2,
\label{potmasshigher}
\end{equation}
where the $a_n$'s are positive or negative coefficients of order unity. As an example,
we calculate $n_s$ and $r$ for $N=60$ in the cases of $a_2=(-0.4) - 0.5$ for 
$K=0.13$ and $a_n=0$ $(n\ge3)$. The results are shown as the green line in the 
left panel of Fig.~\ref{fig2}. The magenta line denotes the same results as that in Fig.~\ref{fig1}, 
and the black dot represents the original quadratic chaotic inflation.
We see that the green line almost coincides with the magenta line. 

In order to confirm that this is a generic feature, we estimate the parameters $n_s$ and $r$ for the 
potential (\ref{potmasshigher}) up to the second and third orders. 
We vary the model parameter $a_2$ from $-1/2$ to 1/2 for 
$K=0.13 - 0.17$ in the former case, while we change $a_3$ from $-1/6$ to $1/6$ for 
$a_2=(-0.2)-0.5$ with $K=0.14 - 0.16$ in the latter case. 
We show the results for corrections up to first (magenta), second (green), and third (blue) 
order in the right panel in Fig.~\ref{fig2}. We find that the derived parameters 
$n_s$ and $r$ reside in a rather narrow region and are very close to the first-order result, 
in spite of including the second- and third-order corrections for various values of the model 
parameters $a_2$ and $a_3$. 
We may therefore conclude that the logarithmic corrections to the mass---regardless of their 
order---make the model consistent with the latest observations such as 
the Planck TT,TE,EE+lowE+lensing+BK14+BAO data \cite{Planck} for appropriate coefficients
$K$ and $a_n$.

\begin{figure}[ht!]
\hspace*{9mm}
\includegraphics[width=100mm]{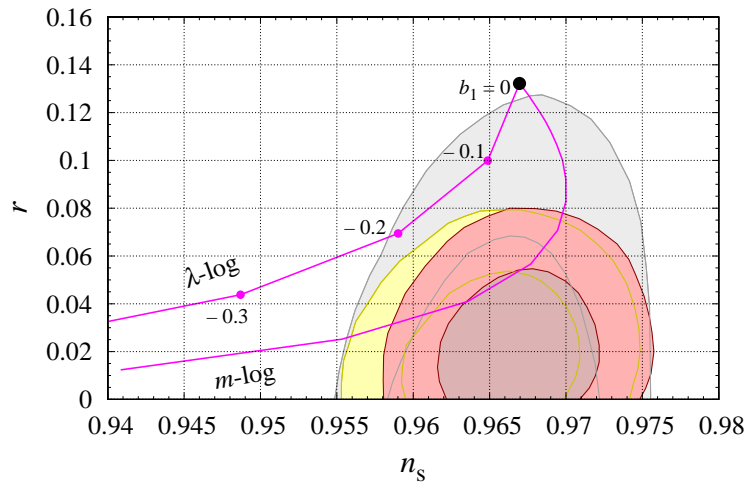} 
\vspace{13mm}
\caption{Spectral index $n_s$ and tensor-to-scalar ratio $r$ for $N=60$ in the case of
the first-order logarithmic corrections to the quartic coupling for $b_1=(-0.4)-0$, 
denoted by the upper magenta line. The lower magenta line denotes the results of the case of 
the first-order corrections to the mass term as in Fig.~\ref{fig1}.
The black dot denotes the original quadratic chaotic inflation case. 
The Planck-allowed regions \cite{Planck} are the same as in Fig.~\ref{fig1}.
\label{fig3}}
\end{figure}

For comparison, we also calculate the case with logarithmic corrections to the 
quartic coupling. The potential is then given by \cite{rad1,rad2,rad3}
\begin{equation}
V=\frac{1}{2} m^2 \phi^2
+\lambda \left[\sum_{n\ge 1}  b_n \left(\log\frac{\phi^2}{M^2}\right)^n \right]\phi^4.
\label{potlambdahigher}
\end{equation}
Here $\lambda \ll 1$ and the $b_n$'s are coefficients of $O((0.1)^n)$. 
We set $b_0=0$, since we consider corrections to the quadratic potential.
For concreteness, we set $\tilde{\lambda}\equiv 2\lambda (M_P/m)^2 = 10^{-3}$ below.

First, we calculate $n_s$ and $r$ for $N=60$ for the first-order correction, 
thus setting $b_n=0$ for $n\ge2$. 
In Fig.~\ref{fig3}, the results of the first-order case for $b_1=(-0.4)-0$ are shown by the magenta 
line (the upper branch, labeled as ``$\lambda$-log"), which is the same as those in 
Refs.~\cite{rad1,rad2,rad3}, while the lower branch (labeled as ``$m$-log") denotes the results 
of the quadratic potential with the first-order correction (\ref{potmass1st}), as in Fig.~\ref{fig1}. 
We see that the two magenta lines are away from each other: the $m$-log corrections 
can be consistent with the Planck TT,TE,EE+lowE+lensing+BK14(+BAO) allowed 
region (red/yellow), while the $\lambda$-log corrections can only explain 
the Planck TT,TE,EE+lowE+lensing data (gray region).

\begin{figure}[ht!]
\hspace*{9mm}
\begin{tabular}{cc}
\includegraphics[width=90mm]{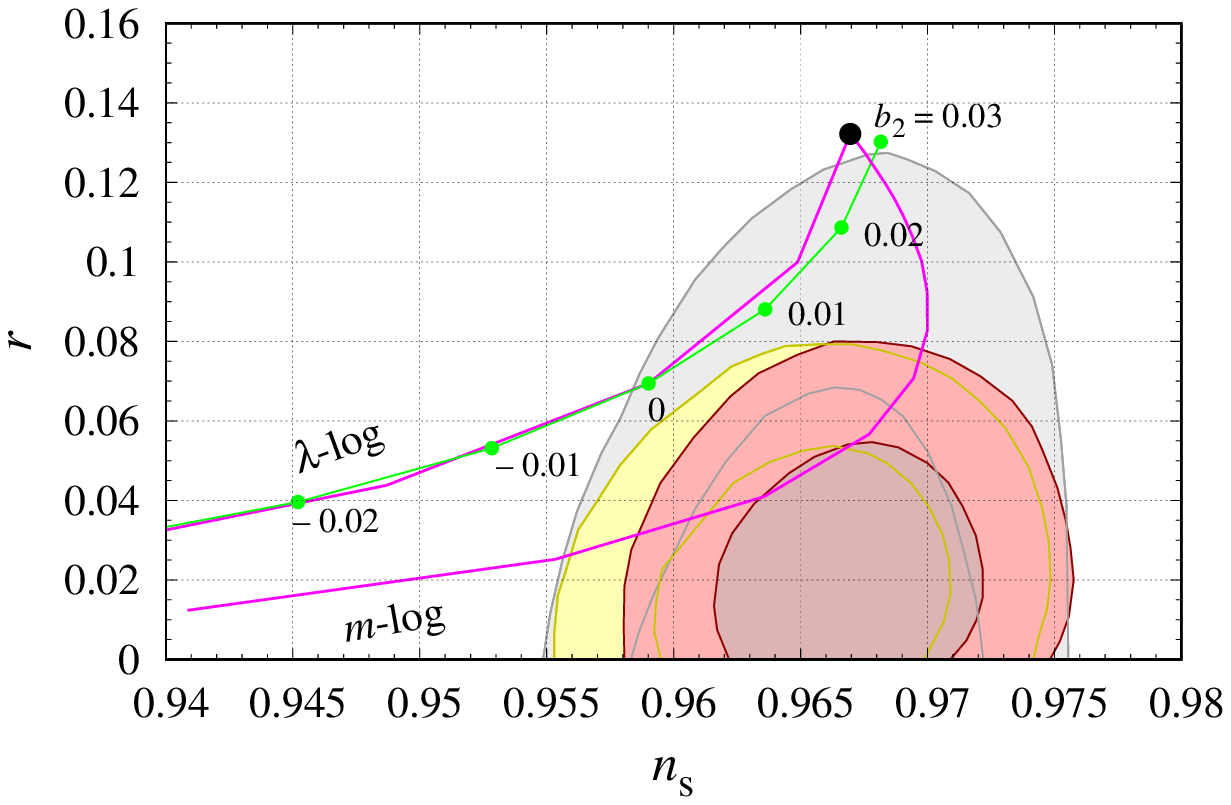} &
\includegraphics[width=90mm]{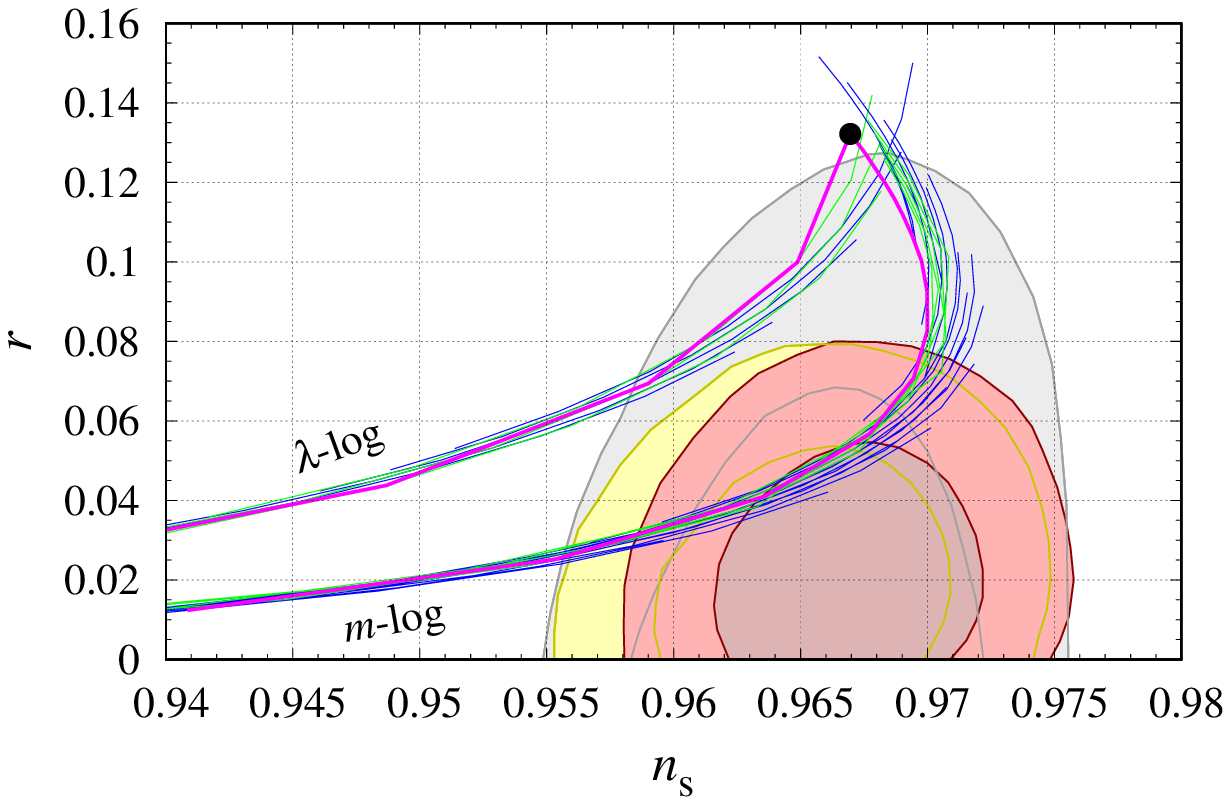} 
\end{tabular}
\vspace{13mm}
\caption{Spectral index $n_s$ and tensor-to-scalar ratio $r$ for $N=60$ in the case of
the higher-order logarithmic corrections to the quartic coupling. The magenta, green, and blue lines 
represent the cases with corrections up to first, second, and third order, respectively.
The black dot denotes the original quadratic chaotic inflation case. 
The Planck-allowed regions \cite{Planck} are the same as in Fig.~\ref{fig1}.
Left: The case for $b_2=(-0.03) - 0.03$ with $b_1=0.2$ and $b_n=0$ $(n\ge3)$, as an example. 
The case of the first-order $m$-log correction is also shown.
Right: All of the cases with both $\lambda$-log and $m$-log corrections explained in the text.
\label{fig4}}
\end{figure}

Let us now include the effects of the second-order corrections. We show the case for 
$b_2=(-0.02)-0.02$ with $b_1=-0.2$ as the green line in the left panel of Fig.~\ref{fig4}. 
We observe that the second-order results are located near the (upper) magenta line.
We also plot the cases for corrections up to the second and third order, respectively, as  
the green and blue lines (the upper branch) in the right panel of Fig.~\ref{fig4}. 
We vary the parameter $b_2$ from $-0.02$ to $0.04$ for $b_1=(-0.3)-(-0.1)$ in the 
former case, while $b_3$ is set between $-0.005$ and $0.003$ for $b_2=(-0.02)-0.02$ 
and $b_1=-0.2$ in the latter case. Again, the resulting parameters $n_s$ and $r$ 
remain in a narrow region close to the (upper) magenta line where only the first order 
correction is included.
We can thus deduce that the model with corrections of the form $\lambda(\log\phi^2)\phi^4$
is only consistent with the Planck data alone, and may already be falsified by
the combinations of the Planck, BICEP2/Keck Array, and/or BAO data.
Therefore, we can distinguish the corrections to the mass from
those to the quartic coupling, so that, in principle, it may reveal the form of the interactions
of the model, when we obtain more precise CMB data in the future.

\section{Conclusions}
We have considered the logarithmic-corrected mass in the chaotic inflation model,
and found that even this simple correction makes the model consistent with observations
such as the CMB observation in {\it Planck} 2018, in combination with 
the BICEP2/Keck Array and baryon acoustic oscillation data.
Since the model predicts a tensor-to-scalar ratio $r$ larger than 0.02 for the present 
allowed values of the spectral index $n_s$, the model will be tested in the near-future 
CMB polarization observations, such as those by the BICEP3/Keck Array \cite{BICEP3}
and the POLARBEAR-2/Simons Array \cite{Simons}.

In addition, we have considered higher-order logarithmic corrections, and found that 
$n_s$ and $r$ stay in a rather narrow region of the parameter space ($n_s$, $r$). 
Moreover, they reside in a completely different region from that for the logarithmic
corrections to the quartic coupling as $\lambda(\log \phi^2) \phi^4$.
Therefore, future CMB polarization observations may figure out what the higher-order
corrections would be, or even single out the form of the interactions.

\section*{Acknowledgments}
S. K. is grateful to Masahiro Kawasaki for helpful conversations.



\end{document}